\documentclass[a4paper,aps,twocolumn,nobibnotes,pra]{revtex4-1}

  \usepackage{ulem}
  \usepackage{color}
  \usepackage{graphicx}
  \usepackage{amssymb}
  \usepackage{amsmath}
  \usepackage{bm}
  \usepackage{verbatim}
  \usepackage[utf8]{inputenc}
  \usepackage{bbm}
  \usepackage{gensymb}
  \usepackage{float}

\newcommand{\ket}[1]{\mbox{$ | #1 \rangle $}}
\newcommand{\bra}[1]{\mbox{$ \langle #1 | $}}
\newcommand{\be}{\begin{eqnarray}}
\newcommand{\ee}{\end{eqnarray}}

\newcommand{\PP}{\ensuremath{\mathcal{P}}}

\begin{document}

\title{Experimental demonstration of robust quantum steering}
\author{Sabine Wollmann$^{1,2}$}
\author{Roope Uola$^{3}$}
\author{Ana C. S. Costa$^{4}$}
\medskip
\affiliation{\textsuperscript{1}Quantum Engineering Technology Labs, H. H. Wills Physics Laboratory and Department of Electrical and Electronic Engineering,
University of Bristol, Bristol BS8 1FD, United Kingdom \\
\textsuperscript{2}Centre for Quantum Computation and Communication Technology (Australian Research Council),
Centre for Quantum Dynamics, Griffith University, Brisbane, Queensland 4111, Australia \\
\textsuperscript{3}D\'epartement de Physique Appliqu\'ee, Universit\'e de Gen\`eve, CH-1211 Gen\`eve, Switzerland\\
\textsuperscript{4}Department of Physics, Federal University of Paran\'a, 81531-980 Curitiba, PR, Brazil}
\date{\today}
\begin{abstract}
We analyse and experimentally demonstrate quantum steering using criteria based on generalised entropies and criteria with minimal assumptions based on the so-called dimension-bounded steering.
Further, we investigate and compare their robustness against experimental imperfections such as misalignment in the shared measurement reference frame.
Whilst entropy based criteria are robust against imperfections in state preparation, we demonstrate an advantage in dimension-bounded steering in the presence of measurement imprecision. As steering with such minimal assumptions is easier to reach than fully non-local correlations, and as our setting requires very little trust in the measurement devices, the results provide a candidate for the costly Bell tests while remaining highly device-independent.

\end{abstract}
\pacs{03.65.Ud, 03.67.-a}

\maketitle

{\it Introduction.---}
Communication protocols based on quantum information have come a long way from abstract theoretical models to everyday technological applications. Some of the most celebrated achievements are undoubtedly the randomness generators~\cite{Law2014} and quantum key distribution~\cite{Acin2006,Branciard2012}. Such protocols demonstrate quantum advantage compared to their classical counterparts by utilizing non-classical resources such as coherence, entanglement, and measurement incompatibility.
Their verification is typically performed in a device-dependent manner, which implies trust in the measurement devices in the laboratory to perform precisely as their manufacturer promises. However, there is no guarantee that these will function exactly as expected and will not be exploited by an adversary. Hence, one would like to reduce the level of trust on the devices. 
The most rigorous way to verify non-classical resources completely trust-free is a Bell test~\cite{bell,bell-review}. Bell tests are fully-device independent and treat measurement devices as black boxes. However, the realization of such tests is experimentally challenging and extremely resource intensive, despite today's technology.
To overcome these difficulties, a relaxation of Bell tests -- quantum steering -- has received a considerable amount of attention~\cite{Wiseman2007,Cavalcanti2007,Uola2019, Bennet2012, Woodhead2016}. \\
Steering based tasks are semi-device independent in that trust is required on one party's, say Bob's, measurement devices while the other party's, say Alice's, devices are treated as black boxes. These protocols show robustness to experimental imperfections and noise, which have to be considered in practical tests~\cite{Wiseman2007,Cavalcanti2007,Uola2019,Weston2018, Wollmann2018}.
Ideally, the remaining trust on Bob's devices is completely removed, however, this requires extremely high-end equipment. Here we demonstrate a steering protocol with minimal trust
by simplifying Bob's devices to the number of degrees of freedom they are monitoring. Further, we consider misalignment to measurement-reference-frames to test for applicability in near-term quantum applications such as quantum fibre networks.  
 \\
In this manuscript, we focus on two promising classes of steering criteria: generalised entropic steering criteria based on Shannon, Tsallis, and R\'enyi entropies~\cite{Schneeloch13,Costa2018,Krivachy2018,footnote} and dimension-bounded steering inequalities~\cite{Moroder2016}. 
Entropic steering criteria allow for the detection of a large class of two-qubit states, can be extended to high dimensional systems, and have been reported to have a detection advantage over linear steering inequalities in terms of noise robustness~\cite{Costa2018}. 
Albeit these advantages, all protocols for quantum steering are based on one party being trusted whilst the other is untrusted. Establishing trust can be very resource intensive, hence protocols making such assumptions redundant are advantageous.  
An example is dimension-bounded steering which allows for the detection of steering from correlations with minimal assumptions about Bob's measurement devices~\cite{Moroder2016}. Whilst assumptions about the Hilbert space dimension that Bob's devices act on remain, none are made about the exact form of Bob's measurements. This brings steering protocols considerably closer to the fully-device independent Bell tests~\cite{Woodhead2016}.
\\
We test both protocols against the misalignment of the shared measurement reference frame and show that the demonstration of quantum steering using dimension-bounded steering is more robust than generalised entropic criteria and allows to overcome previous limitations~\cite{Shadboldt2012, Palsson2012}. Furthermore, we demonstrate the robustness of the dimension-bounded criteria in an even tougher test by considering the scenario where Alice and Bob perform random measurements. Remarkably, for the typical steering scenario of three orthogonal qubit measurements, such randomness does not affect the steerability of the noisy singlet state, i.e. the probability of violation is shown to be $100\%$.
We believe that the results are encouraging for theoretical as well as practical developments in entanglement-based quantum communication protocols beyond scenarios considered here and especially beyond the standard semi-device independent paradigm. This provides an alternative of reaching high-end device-independence with rather low and experimentally friendly fidelities.


{\it Quantum steering.---} 
In a general steering scenario, one assumes that two parties, Alice and Bob, share a quantum state $\varrho_{AB}$. In each round, Bob receives his part of the shared state and announces a randomly chosen measurement setting $x \in \{1,...n\}$ for Alice. Then Alice declares her corresponding measurement outcome $a$ on her system which could be either a fabricated result or a genuine measurement outcome.
Over many runs, Bob can obtain the correlation matrix, which is the joint probability distribution of the measurement outcomes, and test if it can be explained by a local hidden state (LHS) model~\cite{Wiseman2007}. To define a LHS model we consider a state assemblage of Bob's unnormalised states conditioned on Alice's measurement $x$ and outcome $a$ given as $\varrho_{a|x}:=\text{tr}_A[A_{a|x}\otimes\openone\varrho_{AB}]$, where $\{A_{a|x}\}_a$ is a positive-operator valued measure for each $x$, i.e. $\sum_a A_{a|x}=\openone$ and $A_{a|x}\geq0$ for each $a$, and $\varrho_{AB}$ is the state shared between Alice and Bob. The state assemblage allows a LHS model whenever 
\begin{align}
    \varrho_{a|x}=\sum_\lambda p(\lambda)D(a|x,\lambda)\sigma_\lambda,
\end{align}
where $\{p(\lambda)\sigma_\lambda\}_\lambda$ is a state ensemble on Bob's side and $D(\cdot|x,\lambda)$ is a deterministic probability distribution for each $x$ and $\lambda$. If such an LHS model does not exist, Bob can conclude that the shared state $\rho_{AB}$ is entangled. In this way, Alice can steer Bob's system via her measurements. LHS models can be also defined on the level of correlations in which case we say that Alice can steer Bob if the following decomposition of the correlation table $\{p(a,b|x,y)\}$ is not possible
\begin{align}
    p(a,b|x,y)=\sum_\lambda p(\lambda)p(a|x,\lambda)p^Q(b|y,\lambda),
\end{align}
where $p(\cdot|x,\lambda)$ are classical probability distributions and $p^Q(b|y,\lambda)$ refers to a distribution that originates from Bob's measurements on a local state $\sigma_\lambda$. Whenever Bob can perform local tomography, the definitions are equivalent. Here, our criteria are based on correlation tables, but to introduce dimension-bounded steering, we use state assemblages. It should be mentioned that despite using the assemblages in our theoretical considerations, the criteria can be nevertheless evaluated from correlations.

{\it Steering inequalities from general entropic uncertainty relations.---} 
General entropic uncertainty relations provide a state-independent tool to construct steering criteria. Two independent groups~\cite{Costa2018,Krivachy2018} proposed such criteria based on the Tsallis entropy~\cite{Havrda1967,Tsallis1988} and R\'enyi entropy~\cite{Renyi1970}, respectively. The former is parametrised by $q > 1$, and is given by
\be
\label{tsallis}
S_q (\PP) = - \sum_i p_i^q \ln_q (p_i),
\ee
for a general probability distribution $\PP = (p_1,\dots,p_n)$, where the $q$-logarithm is defined as $\ln_q(x) = ({x^{1-q}-1})/({1-q})$.  In the limit of $q \rightarrow 1$, this entropy converges to the Shannon entropy~\cite{coverthomas}.

In the following we consider the case where all outcomes are labelled by $\pm1$, and Bob's measurements correspond to a set of orthogonal spin directions on the Bloch sphere such that $B_m = \mathbf{b}_m \cdot \hat{\sigma}$ with $\mathbf{b}_m \cdot \mathbf{b}_{m'} = \delta_{mm'}$. Here $\mathbf{\hat{\sigma}} = ( \hat{\sigma}_1, \hat{\sigma}_2, \hat{\sigma}_3 )$ is the vector of Pauli operators in some fixed basis.

If the entropy given by Bob's $m$ measurement settings and its $B_m$ outcomes can be bounded by the Tsallis entropic uncertainty (EUR) bound, $\sum_m S_q(B_m)\geq C_B(q,m)$ (for further discussions see Ref.~\cite{Costa2018a}), it is possible to construct steering inequalities in the form of
\begin{equation}
\label{tsc-prob}
\mathcal{S}^{(q)}_m= C_B(q,m) - \frac{1}{q-1}\left[\sum_{m}\left(1 - \sum_{ab}\frac{(p_{ab}^{(m)})^q}{(p_{a}^{(m)})^{q-1}}\right)\right] \leq 0.
\end{equation}
Here, $p_{ab}^{(m)}$ is the probability of Alice and Bob for outcome $(a,b)$ when measuring $A_m\otimes B_m$, and $p_a^{(m)}$ are the marginal outcome probabilities of Alice's measurement~$A_m$. 

If the quantity $S^{(q)}_m$ is positive then the system is steerable.
This form of the steering criteria is not restricted to the case of two-level systems~\cite{Costa2018} and allows for evaluation of any set of measurements, as long as they have a valid entropic uncertainty bound.

Alternatively, generalized entropic steering criteria can be constructed using the R\'enyi entropy~\cite{Renyi1970,Krivachy2018} which is given by
\be
H_r (\PP) = \frac{1}{1-r}\ln\left(\sum_i p_i^r\right).
\ee
In the limit of $r \rightarrow 1$, R\'enyi entropy converges to the Shannon entropy.

The conditional entropies of $A_m$ and $B_m$ can be bounded by a LHS model~\cite{Krivachy2018}, and 
the R\'enyi entropic steering parameter is
\be\label{renyi-parameter}
\mathcal{H}_2^{(r,s)} = R_B(2) - H_r(B|A) - H_s(B|A) \leq 0, \quad
\ee
with the entropy of order $r,s\geq 1/2$, such that $r^{-1}+s^{-1} = 2$. If $r$ and $s$ fulfill these conditions, the bound $R_B(2)$, independent of the order, trivialises to the EUR bound for the Shannon entropy~\cite{MU1988}. Note that Eq.~\eqref{renyi-parameter} only holds for the two-measurement settings scenario, whereas the Tsallis entropic steering criteria do not have such restriction.

{\it Dimension-bounded steering.---} Building dimension-bounded steering tests is a three-step process: First, any unsteerable state assemblage can be prepared with a separable state. For example, consider an unsteerable assemblage $\varrho_{a|x}$ with two inputs, two outputs, and a LHS model given by the operators $\{\omega_{ij}\}$. A separable state that can be used to prepare such assemblage is given by $\sigma_{AB}:=\sum_{i,j}|ij\rangle\langle ij|\otimes\omega_{ij}$ and Alice's corresponding measurements are given by $M_{i|1}:=|i\rangle\langle i|\otimes\openone$ and $M_{j|2}:=\openone\otimes|j\rangle\langle j|$. Second, dimensions-bounded entanglement criteria are accessed by removing the discord zero structure of the states $\sigma_{AB}$ by replacing the operators $|ij\rangle\langle ij|$ with positive operators $Z_{ij}$ and denotes the resulting operator as $\Sigma_{AB}$. Third, one solves as many members of the LHS model as possible using the state assemblage, and eliminates any leftover terms by posing extra constraints for the operators $Z_{ij}$. In our example
\begin{align}
    \Sigma_{AB}:=&\sum_{i,j}Z_{ij}\otimes\omega_{ij}\nonumber\\
    =&Z_{+-}\otimes\varrho_{+|1}+Z_{-+}\otimes\varrho_{+|2}+Z_{--}\otimes\varrho_\Delta\nonumber\\
    +&(Z_{++}-Z_{+-}-Z_{-+}+Z_{--})\otimes\omega_{++},
\end{align}
with $\varrho_\Delta=\varrho_{-|1}-\varrho_{+|2}$ and the elimination of $\omega_{++}$ is done by setting $Z_{++}-Z_{+-}-Z_{-+}+Z_{--}=0$. 

For any unsteerable state assemblage and any set of operators satisfying this elimination criterion, the operator $\Sigma_{AB}$ is a separable quantum state~\cite{Moroder2016}. However, operators corresponding to an entangled state $\Sigma_{AB}$ exist whilst simultaneously satisfying the elimination criterion. This provides a method of mapping steering problems into problems of entanglement detection, for which there exist dimension-bounded techniques.

The entanglement of such states can be witnessed from the steering data in a dimension-bounded manner. The relevant criterion is evaluated through the data matrix $D_{ky}=\text{tr}[G_k\otimes B_y\Sigma_{AB}]$, where $B_y=M_{+|y}-M_{-|y}$ with $M_{\pm|y}$ being Bob's measurement operators and $G_k$ are orthonormal Hermitian operators. The determinant of the data matrix can be used to lower bound the trace norm of a correlation matrix, i.e. a quantity for which an upper bound is known for separable states. This leads to the dimension-bounded steering inequality (for details see~\cite{Moroder2016})
\begin{align}\label{dimbound}
    |\text{det}D|\leq\frac{1}{\sqrt{d_A}}\Big(\frac{\sqrt{2 d_A}-1}{m\sqrt{d_A}}\Big)^{m},
\end{align}
where $m$ is the number of Bob's measurements and $d_A$ is the dimension of the chosen operators $Z_{ij}$. 
In two-qubit systems ($d_A = d_B = 2$), we can define a steering parameter for the dimension-bounded steering by reducing Eq.~(\ref{dimbound}) to
\be\label{DBm}
\mathcal{DB}_m = |\det D| - \frac{1}{\sqrt{2}}\left(\frac{1}{\sqrt{2}m}\right)^{m} \leq 0.
\ee

{\it Entropic and dimension-bounded steering using mutually unbiased bases.---}
In general, protocols for testing or exploiting quantum correlations assume mutually unbiased based (MUB) measurements and a common reference frame between two parties. Their role has recently been investigated using a steering inequality that allows for deterministic violation for a larger class of states~\cite{Wollmann2018}.
Here we implement the same framework for steering criteria based on EURs and dimension-bounded steering, and investigate the robustness of MUB to noise and the role of the number of measurement settings using a subset of data from~\cite{Wollmann2018}. 

\begin{figure}[htpb]
\includegraphics[width=0.42\textwidth]{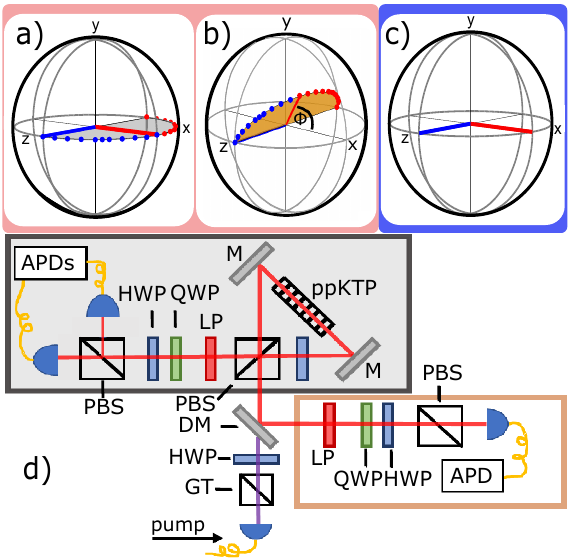}
\caption{The Poincar\'e (Bloch) spheres (a-c) contain vectors showing one of the eigenstates of the two relevant directions (blue and red) in the experiments we performed.   (a) Alice's directions, in the case where Alice and Bob share a reference direction. We test the robustness of our inequalities to rotations in the plane (yellow), as the blue and red settings are rotated through $90{^\circ}$ in steps (blue and red dots). $\Phi=0^{\circ}$ denotes the fact that the plane is not tilted with respect to Bob. (b) Alice's directions for $m=2$ are tilted by $\Phi=30^{\circ}$. (c) Bob uses the same two measurement directions in each experiment. (d) The experiment consisted of entangled photon pair generation at 820 nm via SPDC in a Sagnac interferometer constructed of a polarizing beam splitter (PBS), two mirrors (M), a dual-coated half-wave plate (HWP), and a periodically poled KTP (ppKTP) crystal. Different measurement settings are performed by rotating HWP and quarter-wave plates (QWP) relative to the PBS. Long pass (LP) filters and an additional bandpass filter in Bob's line, remove 410 nm pump photons copropagating with the 820 nm photons before photons are coupled into single-mode fibers and detected by single photon counting modules and counting electronics.}
\label{MUB}
\end{figure}

Here, we consider a two-qubit state, i.e. a Werner state --- $\varrho_\mu = \mu \ket{\psi_s}\bra{\psi_s} + \frac{1-\mu}{4}\mathbbm{1}_4$ ---
which is a probabilistic mixture of a maximally entangled singlet state $\ket{\psi_s}$ with a symmetric noise state parametrized by the mixing probability $\mu \in [0;1]$~\cite{Werner1989}.
Firstly, we limit Bob to fixed MUBs for his measurements (Fig.~\ref{MUB}c), whilst Alice chooses MUBs that can be rotated with respect to the shared reference direction with Bob. Although Alice's and Bob's measurement directions will lie in the same plane, their relative orientation (which we denote as $\alpha$) within this plane may be unknown (Fig.~\ref{MUB}a).

For maximally aligned shared reference directions, the measurements lie in a plane,  
corresponding to an angle of $\Phi = 0\degree$ between Alice's and Bob's measurement planes. Further, we consider the case when Alice and Bob do not share the same reference direction and Alice's reference plane spanned up by her measurements is tilted by $\Phi \neq 0\degree$ (Fig.~\ref{MUB}b).

To verify how these rotations of Alice's MUBs affect the detection of steering, we apply her measurement settings $A_m = \frac{1}{2}(\mathbbm{1} \pm \vec{u}_m\cdot\vec\sigma)$ on the shared state, where $\vec{u}_m \in \mathbb{R}^3$ depends on Alice's measurements orientations ($\alpha$ and $\Phi$) on the Bloch sphere. 

We test our steering protocol considering the case of minimal set size -- $m=2$ MUBs on Alice's side and Bob's side, where we have the bounds $C_B(q,2)= \ln_q (2)$~\cite{Costa2018a} and $R_B(2) = \ln (2)$~\cite{Rastegin2013} for the Tsallis and R\'enyi steering criteria, respectively. Then Eq.~(\ref{tsc-prob}) simplifies to 
\be\label{esc-2}
\mathcal{S}^{(q)}_2 = \frac{1}{(1-q)}\Big[1 + 2^{(1-q)} - f_q(\mu \cos\alpha) \nonumber \\ - f_q(\mu \cos\Phi\cos\alpha)\Big], 
\ee
with $f_y(x):=\left(\frac{1-x}{2}\right)^y + \left(\frac{1+x}{2}\right)^y$, and Eq.~\eqref{renyi-parameter} results in
\be\label{renyi-2}
\mathcal{H}_2^{(r,s)} &=& \ln (2) - \frac{r}{1-r}\ln[f_r(\mu \cos\alpha)]^{1/r} \nonumber\\
&&- \frac{s}{1-s}\ln[f_s(\mu \cos\Phi\cos\alpha)]^{1/s}. 
\ee
The most interesting scenario for the R\'enyi entropic steering criteria is the case of $r=1/2$ and $s=\infty$~\cite{footnote2}, which leads to
\be\label{renyi-min-2}
\mathcal{H}_2^{(1/2,\infty)} = - \ln[1+\sqrt{1-\mu^2\cos^2\alpha}] \nonumber \\
 +\ln\big[1+\mu|\cos\Phi\cos\alpha|\big].
\ee
For this class of states and set of measurements, $\mathcal{S}^{(2)}_2$ detects steerability for the same range of parameters of $\mathcal{H}_2^{(1/2,\infty)}$, i.e. both steering parameters are positive if $\mu > 1/[\cos\alpha \sqrt{1+\cos^2\Phi}]$. This shows the equivalence of both criteria in this case.
Furthermore, the constraints on Bob's side reduce Eq.~(\ref{DBm}) to 
\be\label{db-2}
\mathcal{DB}_2 = \frac{1}{8\sqrt{2}}(2\mu^2 |\cos\Phi|-1).
\ee

The steering protocol based on Eqs.~\eqref{esc-2}-\eqref{db-2} is dependent on $\Phi$ and therefore rotationally variant in the case of two MUBs per site. Further, the entropic criteria are limited to specific misalignment ($\alpha$) within the measurement plane. Whilst deviations of Alice's measurement directions will affect the detection of steering, we will show the robustness of entropic criteria for some specific cases and compare it to the dimension-bounded criterion.

{\it Experimental details and results.---}
We implemented the steering protocols using a high-efficiency spontaneous parametric down-conversion (SPDC) source (Fig.~\ref{MUB}d). This source, mounted in a Sagnac Ring interferometer~\cite{Kim2006,Fedrizzi2007}, consists of a 10 mm-long periodically poled potassium titanyl phosphate (ppKTP) crystal pumped bidirectionally by a 410 nm fiber-coupled continuous-wave laser with an output power (after fiber) of 2.5 mW.
The generated state is verified using quantum state tomography~\cite{White2007} at several stages throughout the experiment---in each case, we achieved a fidelity of ca.~$98\%$ with the singlet state $|\Psi^-\rangle$. Alice and Bob's $m$ measurement directions and projective measurements are implemented by rotating the QWPs and HWPs in front of polarising elements together with coincidence detections. The steering parameter and its error are calculated from the observed correlations. The error $\Delta S^{(q)}=\sqrt{(\Delta S_{\textrm{sys}})^2+(\Delta S_{\textrm{stat}})^2}$ consists of a systematic and a statistical error due to small imperfections in Bob's measurement settings and Poissonian statistics in photon counting, respectively.

We investigated the steering protocol using two MUBs aligned along $\sigma_z$ and $\sigma_x$ on Alice's and Bob's side (Fig.~\ref{MUB}(a,c) (blue and red)).
We successfully violated the steering inequalities (Eqs.~(\ref{esc-2}), (\ref{renyi-min-2}) and (\ref{db-2})) with $\mathcal{S}^{(1)~exp}_2= 0.524 \pm 0.008$ (criterion based on the Shannon entropy), $\mathcal{S}^{(2)~exp}_2 = 0.433 \pm 0.004$, $\mathcal{H}^{(1/2,\infty)~exp}_2 = 0.486 \pm 0.008$ and $\mathcal{DB}_2 = 0.076 \pm 0.002$. Note that although the entropic steering criteria allow for stronger violation of the classical bound than the dimension-bounded criteria, the amounts of violation are not comparable with one another without the appropriate normalization.

\begin{figure}[htbp]
\includegraphics[width=0.44\textwidth]{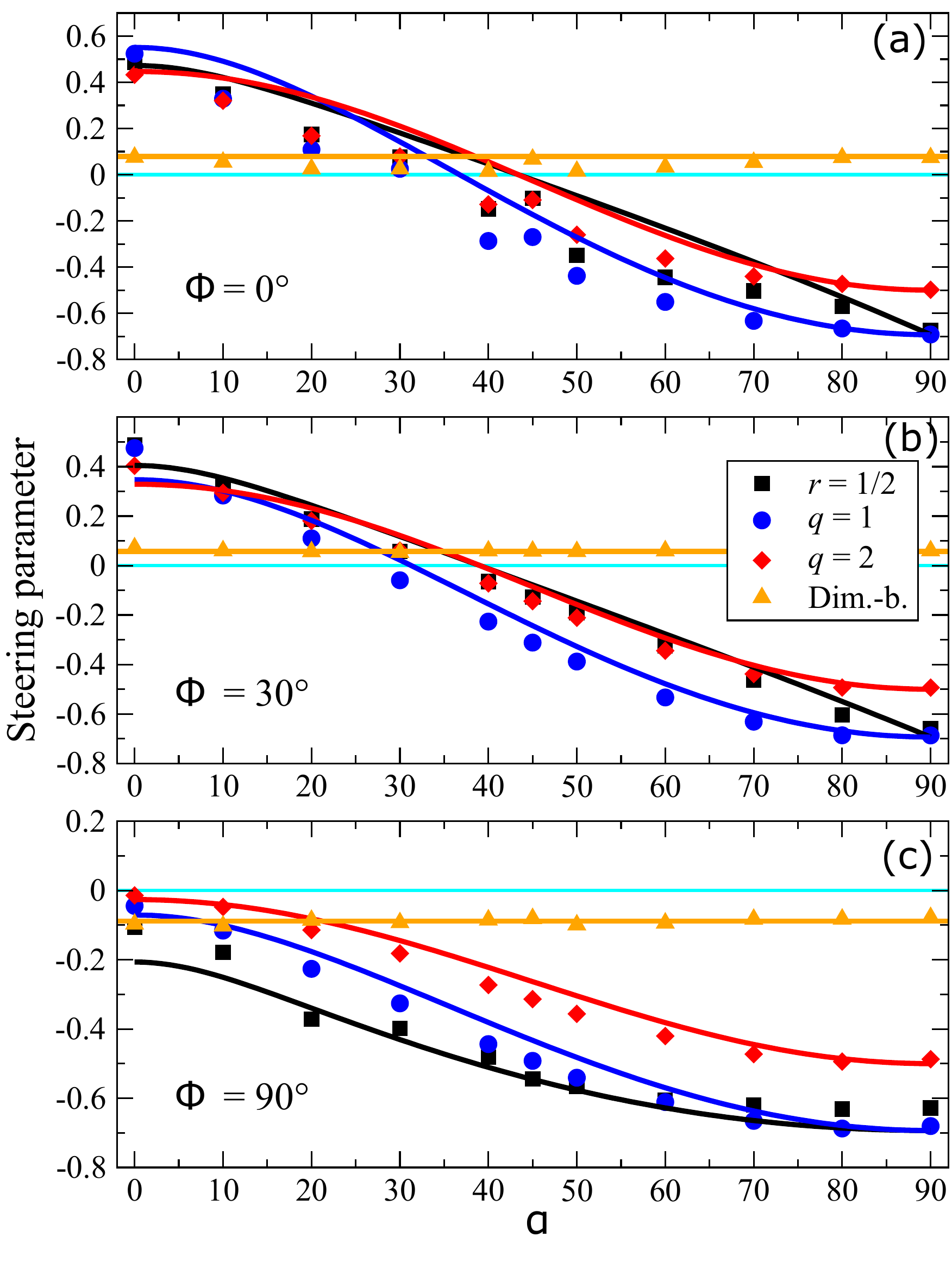}
\caption{Steering parameters for experiments with $m = 2$ measurement directions. Renyi (black square), Shannon (blue circle), Tsallis (red diamond) entropy criteria and dimension bounded steering parameter (yellow triangle) as function of the rotation angle $ \alpha$ (degrees) in Alice's measurement plane. Angle $\Phi$ denotes the angle of tilt between  Alice's and Bob's measurement plane. Error bars are too small to be seen.}
\label{2MUB}
\end{figure}

To test the robustness of Eqs.~(\ref{esc-2}), (\ref{renyi-min-2}) and (\ref{db-2}) we considered misalignment of Alice's and Bob's MUBs by $\alpha$ and $\Phi$. This accounts for realistic situations in a laboratory environment when a shared reference direction may be determined reliably, however, the relative orientation of the observer's MUBs within this plane may be unknown e.g. in quantum networks.

First, they share a single reference direction ($\Phi=0^{\circ}$) and the measurement directions lie in a plane orthogonal (on the Bloch sphere) to the shared direction but are misaligned by $\alpha$ (Fig.~\ref{MUB}a). While Bob's measurement directions were kept constant, Alice's were rotated through $90^{\circ}$ in the plane, by angles
$\alpha \in \{0^{\circ}, 10^{\circ}, 20^{\circ}, 30^{\circ}, 40^{\circ}, 45^{\circ}, 50^{\circ}, 60^{\circ}, 70^{\circ}, 80^{\circ}, 90^{\circ}\}$. 
 
The steering inequalities are robust to misalignment for values of up to $\alpha<36.7^{\circ}$ for the Shannon and $\alpha<43.4^{\circ}$ for the Tsallis and R\'enyi entropy, while the dimension-bounded criterion demonstrates steering for all values of $\alpha$ (Fig.~\ref{2MUB}a). 

Next we increased the misalignment by tilting to angle $\Phi=30^{\circ}$ whilst maintaining MUBs (Fig.~\ref{MUB}b). Our experimental results demonstrate steering for $\mathcal{S}^{(1)}$, $\mathcal{S}^{(2)}$, and $\mathcal{H}^{(1/2,\infty)}$ for $\alpha<31.2^{\circ}$ and $\alpha<39^{\circ}$, respectively (there is no difference between $\mathcal{S}^{(2)}$ and $\mathcal{H}^{(1/2,\infty)}$ in this scenario).

Finally, we investigated the case of no shared reference direction between Alice and Bob ($\Phi=90^{\circ}$). Although they maintain their MUBs for measurement on each side, the lack of reference makes it impossible for Alice to steer Bob's state even for the dimension-bounded steering criterion.
Our investigation shows that in the presence of misalignment entropic steering criteria lose their advantage over dimension-bounded steering. Thus, a detection method with fewer assumptions performs better.

Further, we extend the protocol to three MUBs per site and discuss the details in the Appendix.
The robustness of the Tsallis entropic criteria is improved when considering a triad of measurements for each party, while the dimension-bounded steering becomes completely rotationally invariant.  We provide a detailed analysis of these aspects in the Appendix.

{\it Conclusions.---} 
We have experimentally demonstrated quantum steering using generalized entropic criteria and dimension-bounded steering inequalities and discussed their robustness to reference-frame misalignment. For two and three measurement settings per side, we showed that the criteria can be violated using a sufficiently entangled state. The criteria show robustness to misalignment of the measurement directions with dimension-bounded criteria being more robust of the two. In Ref.~\cite{Wollmann2018}, reference-frame invariance was demonstrated for linear criteria with comparable robustness to that of dimension-bounded steering. However, the latter requires fewer assumptions on Bob's measurements, hence, making it more desirable for future semi-device independent quantum communication protocols. 
Most importantly, this result suggests that near term quantum devices can be based on steering with a very high amount of device-independency.

Further, we demonstrate the equivalence of entropic criteria based on Tsallis and R\'enyi entropies for specific cases.
The Tsallis criteria are preferable as they allow the protocol to be extended to three measurement settings per side, providing greater noise robustness.

An interesting future avenue is to experimentally investigate the steerability of higher dimensional states via generalized entropic criteria and dimension-bounded steering. Moreover, experiments involving multipartite entropic steering~\cite{Cavalcanti2015,Riccardi2018,Costa2018a} seems to be a promising case of interest.

{\it Acknowledgments.---}
We thank Paul Skrzypczyk, Jonathan Matthews, and John Rarity for useful discussions.
S.W. acknowledges financial support by the Australian Government, from EPSRC (EP/M024385/1), and travel support from Q-turn. Further we thank the organisers of the `Q-Turn: changing paradigms in quantum science' workshop for creating stimulating discussions between the authors. A.C.S.C. acknowledges support by the CNPq/Brazil (process number: 153436/2018-2) and CAPES/Brazil. R.U. is grateful for the financial support from the Finnish Cultural Foundation.

{\sl Note added.} -- While finishing this manuscript, we became aware of the work~\cite{Yang2019}, where the generalized entropic steering criteria were experimentally tested in a photonic setup, and the work~\cite{Zhao2019}, on the topic of reducing trust in a steering experiment.

\section*{Appendix}

\subsection*{A: Three-measurement settings for each party}

Here, we consider the case where Alice and Bob each choose a triad of mutually orthogonal directions, i.e., $m = 3$. In this scenario, we focus on the Tsallis entropic steering criteria and the dimension-bounded steering as the R\'enyi entropic steering criteria cannot be implemented for more than two measurements per site. Then the entropic steering inequality is given by
\be\label{esc-3}
\mathcal{S}^{(q)}_3 &=&\frac{1}{(1-q)}\Big[1 + 2^{(2-q)}- f_q(\mu \cos\Phi) - f_q(\mu \cos\alpha) \nonumber \\
&&- f_q(\mu \cos\Phi\cos\alpha)\Big] \leq 0,
\ee
with Tsallis EUR bounds given by $C_B(q,3) = 2\ln_q(2)$, while the dimension-bounded steering is
\be\label{db-3}
\mathcal{DB}_3 = \frac{1}{12}\left(\frac{\mu^3}{\sqrt{3}} - \frac{1}{9}\right) \leq 0.
\ee

\begin{figure}[htbp]
\includegraphics[width=0.46\textwidth]{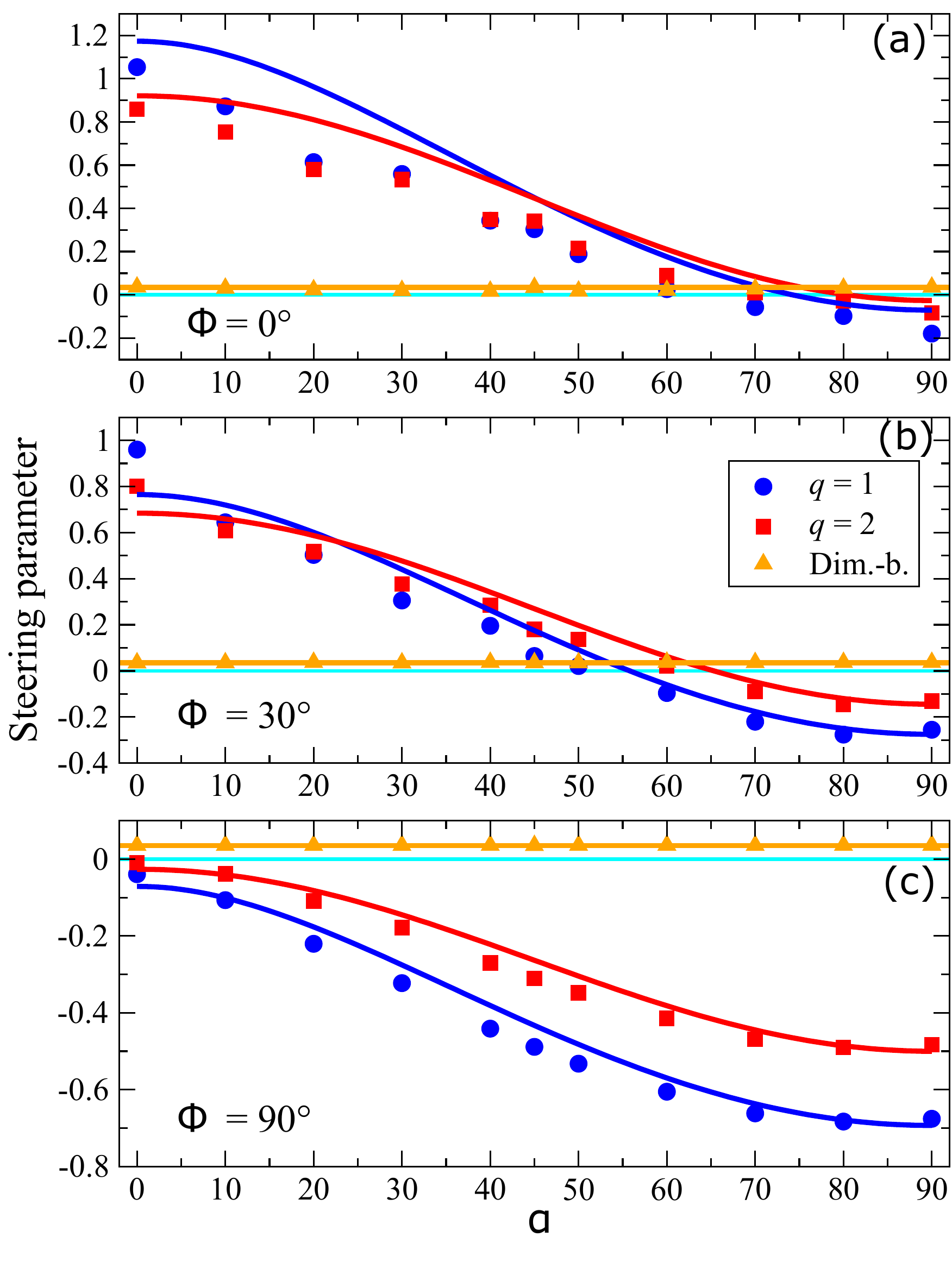}
\caption{Steering parameters for experiments with $m = 3$ measurement directions. Shannon (blue circle), Tsallis (red diamond) entropy criteria and dimension bounded steering parameter (yellow triangle) as function of the rotation angle $ \alpha$ (degrees) in Alice's measurement plane. Angle $\Phi$ denotes the angle of tilt between  Alice's and Bob's measurement plane. Error bars are too small to be seen.}
\label{3MUB}
\end{figure}

Further, we considered the same misalignment scenarios by $\alpha$ and $\Phi$ (Fig.~\ref{3MUB}). First, Alice's and Bob's MUBs lie within the same measurement plane ($\Phi=0^{\circ}$) and are rotated by angle $\alpha$. Whilst the Shannon inequality for $q \rightarrow 1$ (Eq.~(\ref{esc-3})) provides a greater violation of the bound in such scenario at $\alpha=0^{\circ}$, the inequality for $q=2$ allows for demonstration of steering for higher values of $\alpha$ than the Shannon inequality, with $\alpha<80^{\circ}$ and $\alpha<75^{\circ}$, respectively. 
For $\Phi=30^{\circ}$ the inequality for $q=2$ can be violated for $\alpha<66^{\circ}$ whilst the $q \rightarrow 1$ inequality can only tolerate $\alpha<56^{\circ}$. A complete loss of reference direction ($\Phi=90^{\circ}$) does not allow for any demonstration of steering using three MUBs (Fig.~\ref{3MUB}c). Interestingly, dimension-bounded steering for three-measurements per site is rotationally invariant, as one can see in Eq.~\eqref{db-3}, and detects steering in all scenarios.

\subsection*{B: Loss of MUBs}
Further, we investigated the scenario where Alice chooses to perform non-orthogonal measurements (NOM), while Bob's measurements remain orthogonal. 

First, lets consider the case of $m=2$ NOM measurement settings. Bob measures along $\sigma_z$ and $\sigma_x$ and Alice measures along $\vec{u}_1 = (0,0,1)$ and $\vec{u}_2 = (\sqrt{3}/2,0,1/2)$.
This leads to the steering criteria
\be
\mathcal{S}^{(q)}_{2-NOM} &=& \frac{1}{(1-q)}\left[1 + 2^{(1-q)} - f_q(\mu) - f_q\left(\frac{\sqrt{3}\mu}{2}\right)\right], \nonumber \\
\mathcal{H}^{(1/2,\infty)}_{2-NOM} &=& \ln\left[1+\frac{\sqrt{3}}{2}\mu\right] - \ln[1+\sqrt{1-\mu^2}], \nonumber \\
\mathcal{DB}_{2-NOM} &=& \frac{1}{8\sqrt{2}}(\sqrt{3}\mu^2-1).
\ee
The engineered state had a fidelity of 97.2\% with the singlet state, corresponding to the Werner state with $\mu = 0.963$, which allows to demonstrate steering with $\mathcal{S}^{(1)~exp}_{2-NOM}=0.304 \pm 0.016$ and $\mathcal{S}^{(2)~exp}_{2-NOM}=0.316 \pm 0.01$, which strongly agrees with the theoretically determined steering values of $\mathcal{S}^{(1)~theo}_{2-NOM}= 0.314$ and $\mathcal{S}^{(2)~theo}_{2-NOM}= 0.311$, respectively. Concerning the R\'enyi entropic steering criteria and dimension-bounded criteria, the same agreement between the analytical and experimental results can be verified, where we have $\mathcal{H}^{(1/2,\infty)~exp}_{2-NOM}=0.365 \pm 0.01$ and $\mathcal{H}^{(1/2,\infty)~theo}_{2-NOM}=0.367$, and $\mathcal{DB}_{2-NOM}^{~exp}=0.051 \pm 0.004$ and $\mathcal{DB}_{2-NOM}^{~theo}=0.053$.

Finally, we extend the analysis to the case of three measurements. 
While Bob measures along $\sigma_z$, $\sigma_y$ and $\sigma_x$, Alice measures along $\vec{u}_1 = (0,0,1)$, $\vec{u}_2 = (\sqrt{3}/2,0,1/2)$, and $\vec{u}_3 = (1/(2\sqrt{3}),\sqrt{2/3},1/2)$.
This leads to the steering parameters
\be
\mathcal{S}^{(q)}_{3-NOM} &=& \frac{1}{1-q}\Bigg[1 + 2^{(2-q)} - f_q(\mu) - f_q\left(\sqrt{\frac{2}{3}}\mu\right) \nonumber \\
&&- f_q\left(\frac{\sqrt{3}\mu}{2}\right)\Bigg], \nonumber \\
\mathcal{DB}_{3-NOM} &=& \frac{1}{12}\left(\frac{\mu^3}{\sqrt{6}}- \frac{1}{9}\right).
\ee
Our generated state with $\mu = 0.963$ is steerable with $\mathcal{S}^{(1)~exp}_{3-NOM}=0.593 \pm 0.03$, $\mathcal{S}^{(2)~exp}_{3-NOM}=0.544 \pm 0.02$, and $\mathcal{DB}_{3-NOM}^{~exp}=0.021 \pm 0.003$, which agrees well with $\mathcal{S}^{(1)~theo}_{3-NOM}= 0.667$, $\mathcal{S}^{(2)~theo}_{3-NOM}= 0.620$, and $\mathcal{DB}_{3-NOM}^{~theo}=0.021$ for the Shannon and Tsallis entropic steering inequality and the dimension-bounded steering, respectively.

We showed that our steering criteria allow for demonstration of quantum steering when using NOM, the most rigorous form of misalignment. Whilst the violation of the bounds is not very strong, it still emphasizes the suitability of the investigated criteria for future applications in e.g. quantum networks.

\subsection*{C: Probability of demonstrating dimension-bounded steering with random measurements}

Here we extend our analysis to the case where Alice and Bob perform random measurements. We first discuss the probability of violation of the dimension-bounded criterion in the case of random orthogonal measurements, and then we check the case of completely random measurements. 

In the case of two measurement settings and the Werner state, we can rewrite Eq.~(\ref{dimbound}) as
\be\label{2meas-vec}
\mu^2 |(\vec{a}_1\times\vec{a}_2)\cdot (\vec{b}_1\times\vec{b}_2)| \leq \frac{1}{2}.
\ee
Note that in the case that both Alice and Bob perform MUBs, i.e. $(\vec{a}_1,\vec{a}_2)$ (and $(\vec{b}_1,\vec{b}_2)$) are orthogonal, the above inequality becomes $\mu^2|\cos(\gamma)| > \frac{1}{2}$, where $\gamma$ is the angle between the perpendicular vectors to the planes defined by $(\vec{a}_1,\vec{a}_2)$ and $(\vec{b}_1,\vec{b}_2)$ - this result is equivalent to the quantity presented in Eq.~(\ref{db-2}). Therefore, it is straightforward to show that, for a singlet state ($\mu=1$), the probability of violation is 66.7\%.

In the case of three measurement settings, the inequality can be rewritten as
\be\label{3meas-vec}
\mu^3 |\vec{a}_1\cdot (\vec{a}_2\times\vec{a}_3)||\vec{b}_1\cdot (\vec{b}_2\times\vec{b}_3)| \leq \frac{\sqrt{3}}{9}.
\ee
If $(\vec{a}_1,\vec{a}_2,\vec{a}_3)$ (and $(\vec{b}_1,\vec{b}_2,\vec{b}_3)$) are orthogonal, we have that $\mu^3 \leq \frac{\sqrt{3}}{9}$, which is equivalent to the quantity given in Eq.~(\ref{db-3}). Here, one can see that for $\mu > \frac{1}{\sqrt{3}}$, the probability of violation is 100\% - the inequality is violated for all triads of orthogonal measurements. 

\begin{figure}[htbp]
\includegraphics[scale=0.5]{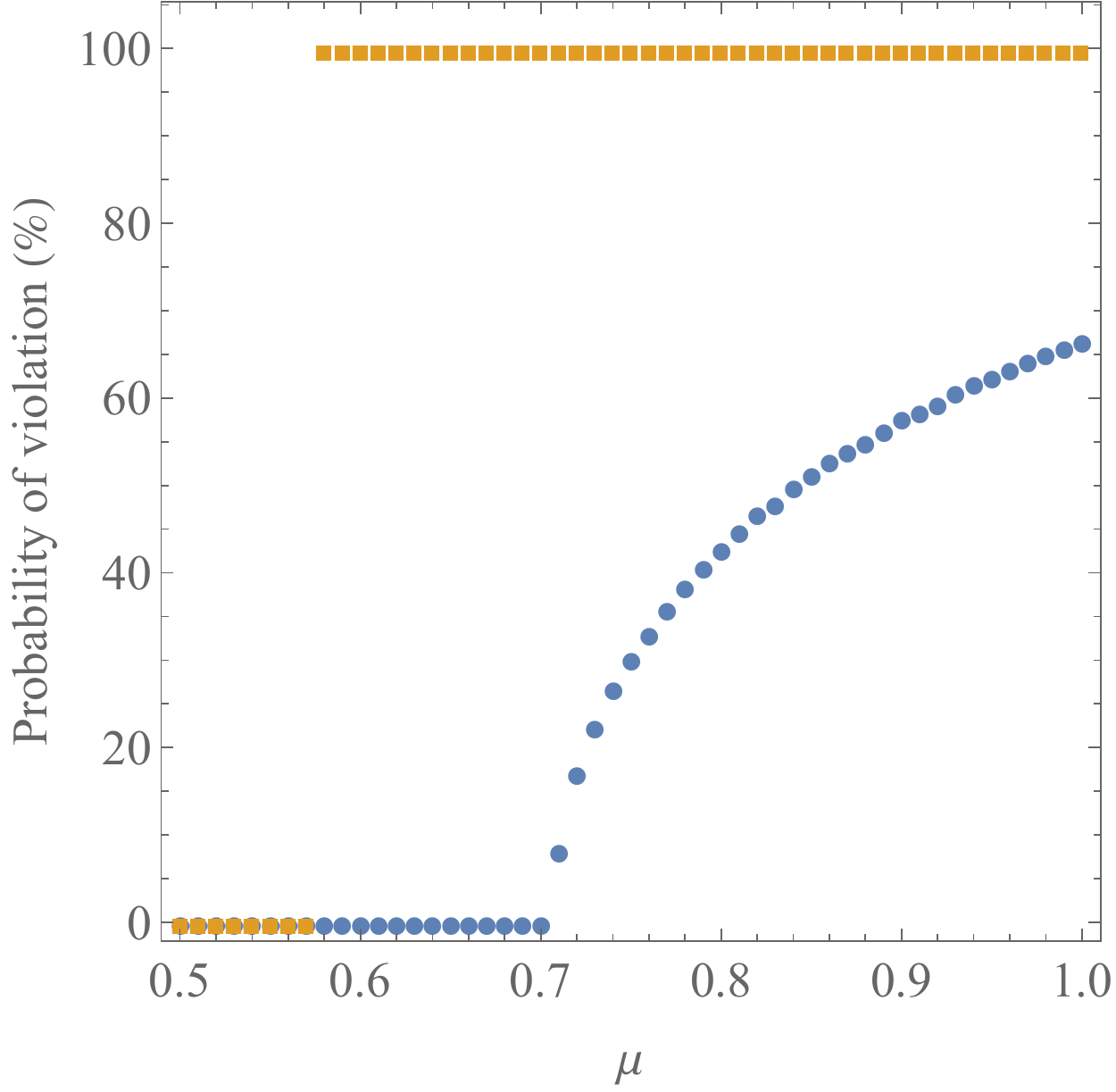}
\caption{Probability of violation as a function of the mixing probability $\mu$, in the case of two (blue circles) and three (orange squares) random orthogonal measurement settings.}
\label{fig2}
\end{figure}
\begin{figure}[htbp]
\includegraphics[scale=0.48]{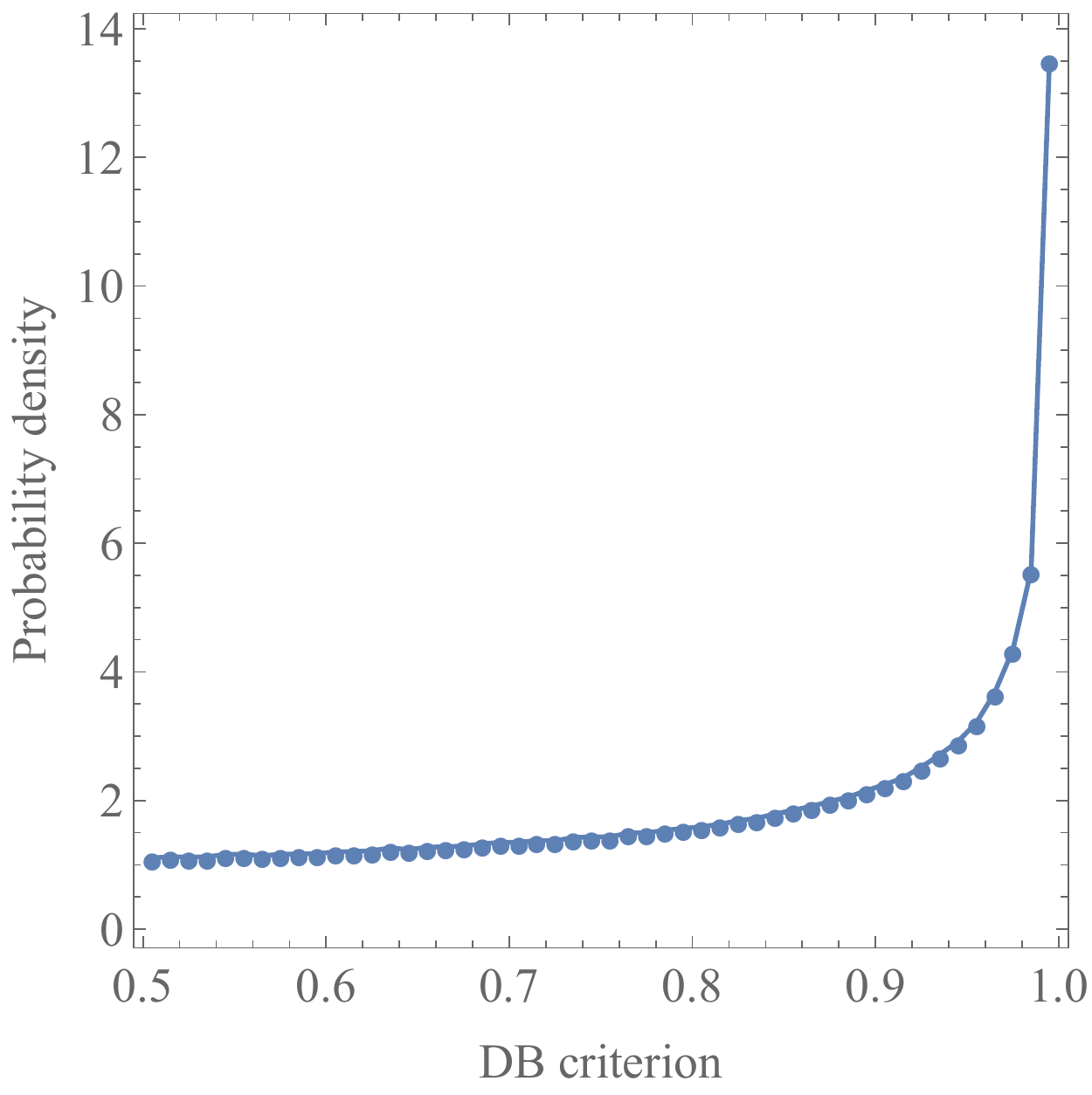}
\caption{Probability density for the amount of violation of the dimension-bounded criterion Eq.~(\ref{2meas-vec}) for a singlet state and two random orthogonal measurements per site.}
\label{fig3}
\end{figure}

In Fig.~\ref{fig2}, we plot the probability of violation as a function of the mixing parameter $\mu$ in the case of random orthogonal measurements performed by Alice and Bob. As discussed before, in the case of two measurement settings the probability increases as a function of $\mu$, with the probability of $66.7\%$ being reached for the perfect singlet, whereas the probability of violation in the case of three measurement is always 100$\%$ for $\mu>1/\sqrt{3}$. Another interesting aspect is the analysis of the probability density as a function of the quantity by which the inequality in Eq.~(\ref{2meas-vec}) is violated. In Fig.~\ref{fig3}, one sees that in the case of two orthogonal measurements, the inequality is mostly violated in its maximum value.

\begin{figure}[htbp]
\includegraphics[scale=0.5]{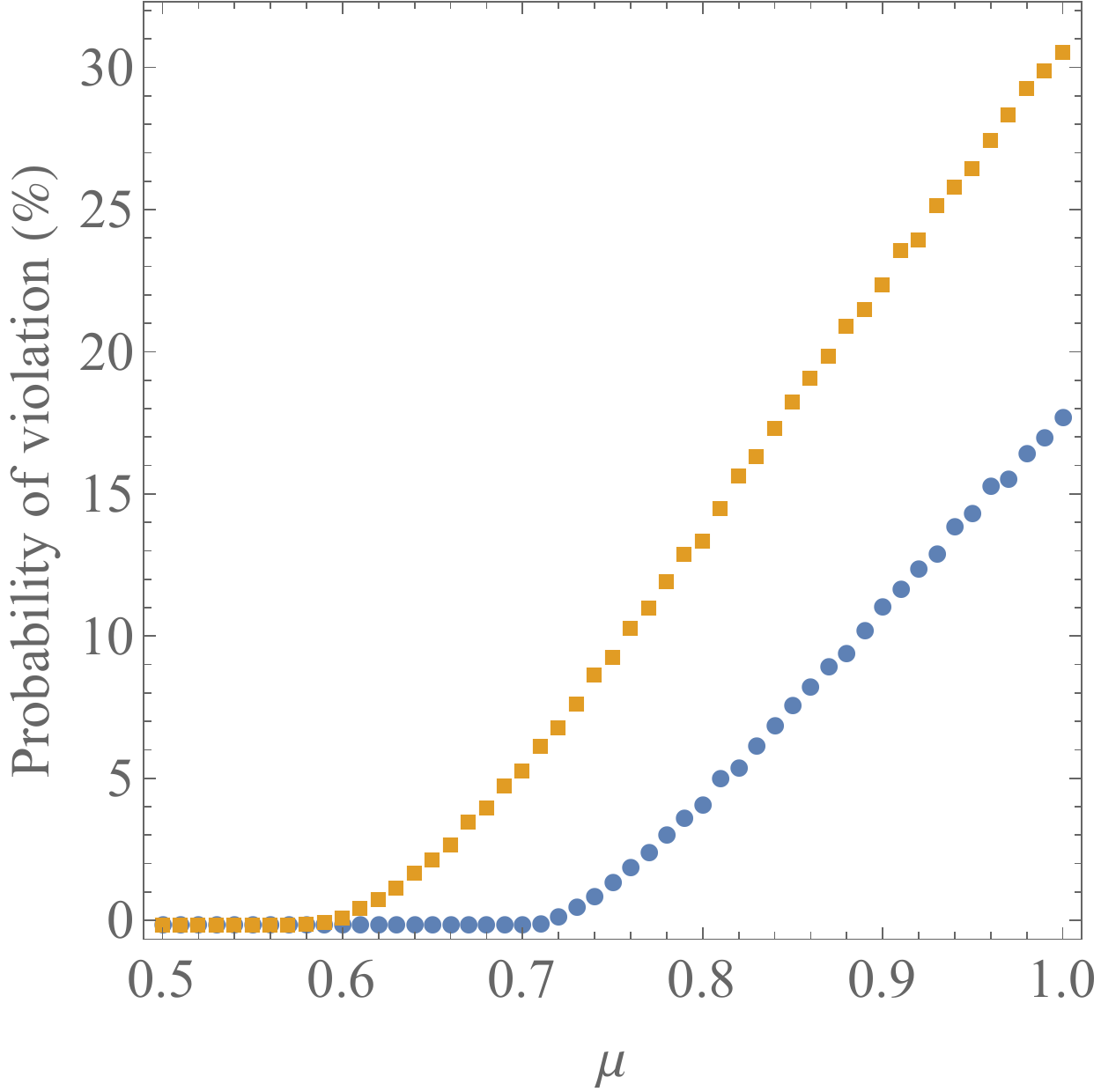}
\caption{Probability of violation as a function of the mixing probability $\mu$, in the case of two (blue circles) and three (orange squares) completely random measurement settings.}
\label{fig4}
\end{figure}

\begin{figure}[htbp]
\includegraphics[scale=0.33]{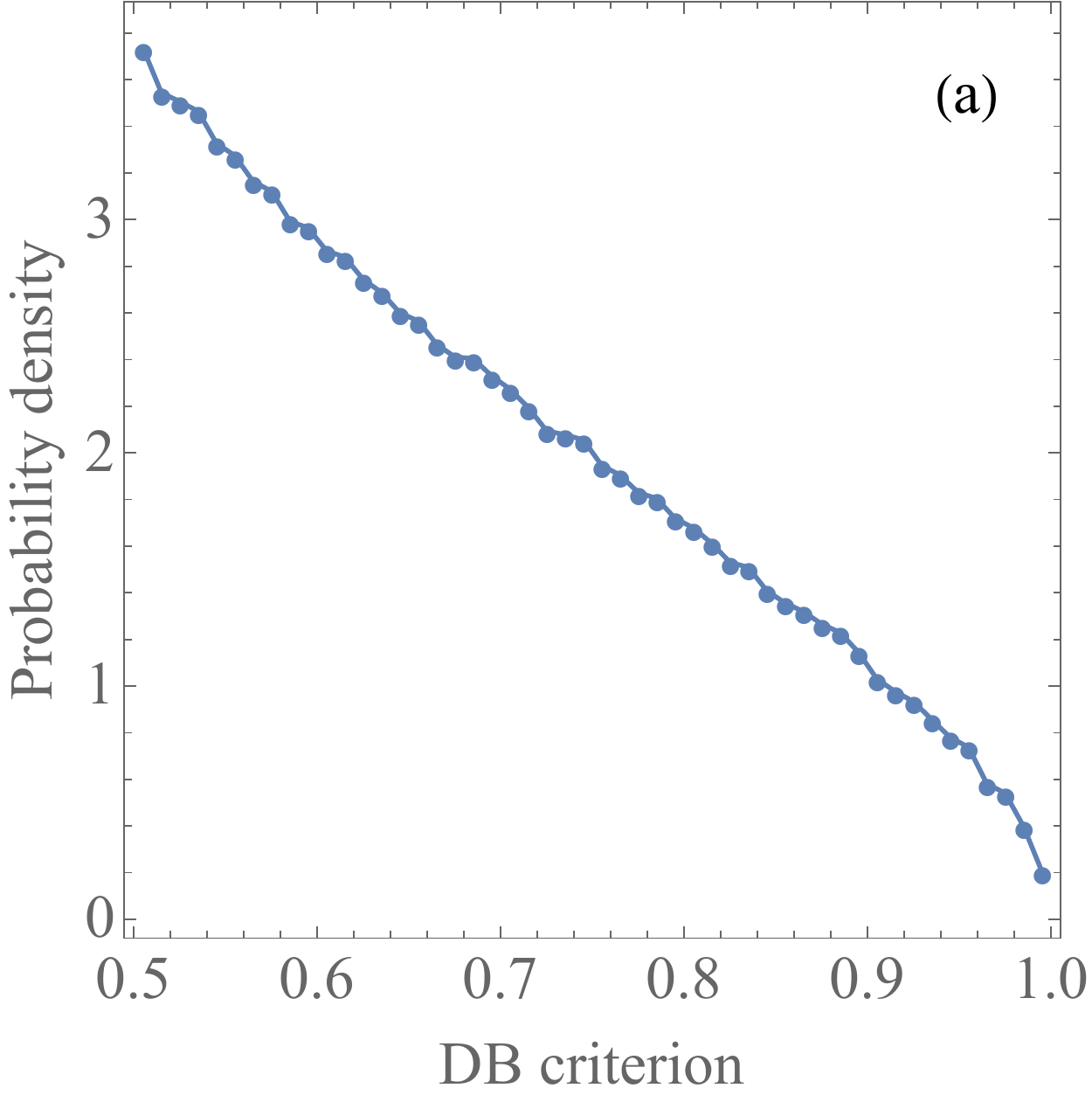}
\includegraphics[scale=0.33]{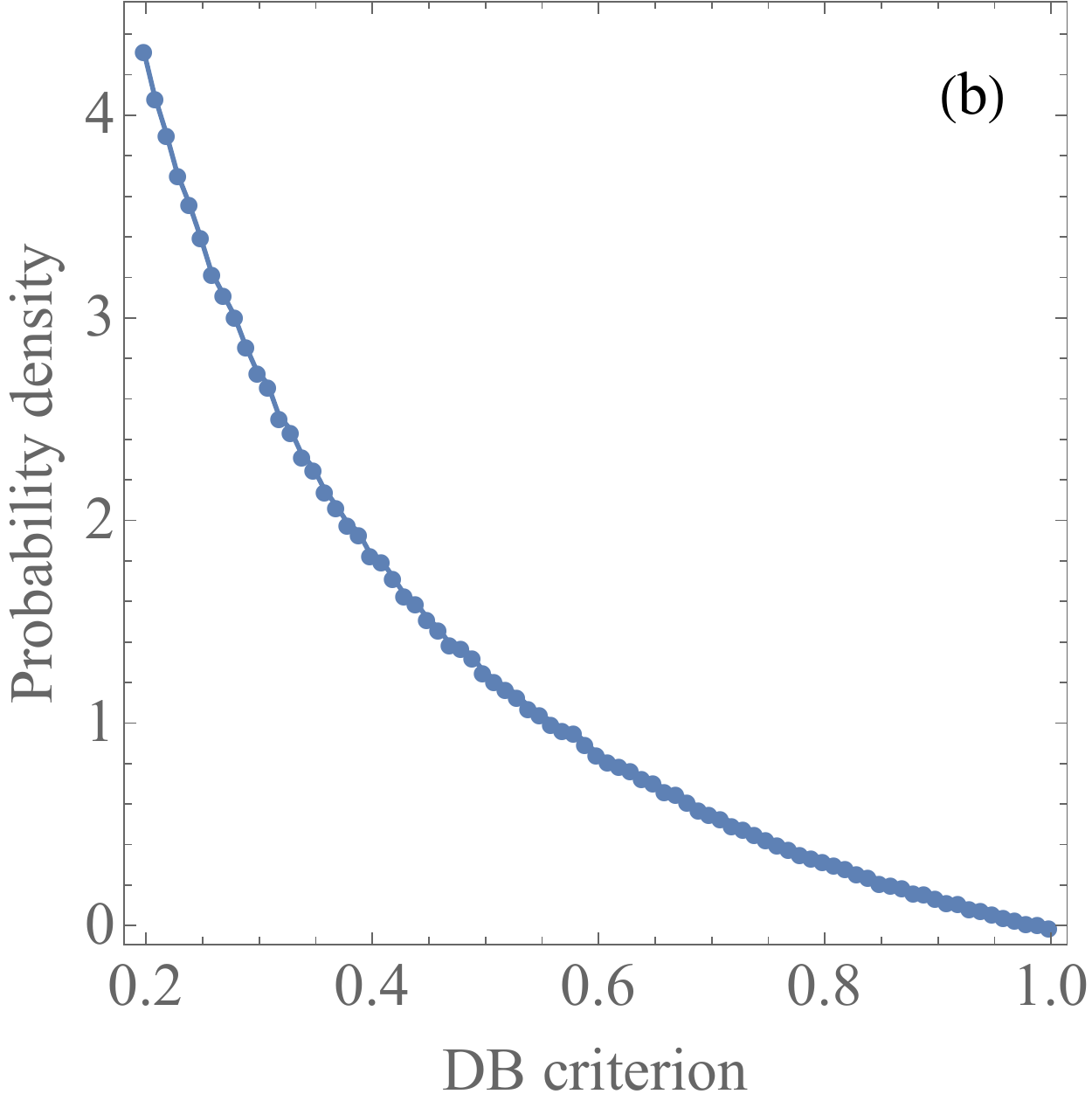}
\caption{Probability density for the amount of violation of the dimension-bounded criterion in (a) Eq.~(\ref{2meas-vec}) by two completely random measurements per site and (b) Eq.~(\ref{3meas-vec}) by three completely random measurements per site for a singlet state.}
\label{fig5}
\end{figure}

In the case that Alice's and Bob's measurements are completely random, i.e. not necessarily orthogonal measurements, the probability of violation naturally decreases. The probability of violation for a perfect singlet state is approximately 17.8\% in the case of two measurement settings and 30.8\% in the case of three measurement settings. In Fig~\ref{fig4}, the probability of violation is demonstrated as a function of the mixing parameter $\mu$, and the probability density for the amount of violation is depicted in Fig~\ref{fig5}.
 
As the final analysis, we check the effect of raising the classical bound on the probability of violation (Tbl.~\ref{table1}) to
take the uncertainty of finite-sized data sets into account \cite{Shadboldt2012}. Our analysis shows, even in the event of
significant estimation imprecision of up to 20\%, statistically significant violations of the classical bound and these
are not concentrated around the local bound when using random measurements. More precisely, when the bound is increased by 10\%, the probability of violation of a singlet state decreases to 62.9\% for two random orthogonal measurements, 14.7\% for two completely random measurements and 28.4\% for three completely random measurements. For an increase of 20\%, the probability of violation for the singlet state decreases further to 59.1\% for two random orthogonal measurements, to 11.9\% for two completely random measurements and to 26.2\% for three completely random measurements. Remarkably, for three random orthogonal measurements, the probability of violation remains in 100\%.

\begin{table}[h]
\begin{tabular}{|c|c|c|c|}
\hline
Increase of the bound by & 0\% & 10\% & 20\% \\ \hline
2 ROM & 66.7\% & 62.9\% & 59.1\% \\ \hline
3 ROM & 100\% & 100\% & 100\% \\ \hline
2 CRM & 17.8\% & 14.7\% & 11.9\% \\ \hline
3 CRM & 30.8\% & 28.4\% & 26.2\% \\ \hline
\end{tabular}
\caption{Displacement of the probability of violation by 10\% and 20\% increase of the dimension-bounded classical bound for a singlet state. Here, we considered random orthogonal measurements (ROM) and completely random measurements (CRM).}
\label{table1}
\end{table}



\begin{thebibliography}{30}

\bibitem{Law2014}
 {{Y. Z.} {Law}},
 {{L. P.} {Thinh}}, 
 {{J.-D.} {Bancal}}, {and}
 {{V.} {Scarani}},
{J. Phys. A} \textbf{{47}}, {424028} ({2014}).

 \bibitem{Acin2006}
 {{A.} {Acin}},
 {{N.} {Gisin}}, {and}
 {{L.} {Masanes}}, 
{Phys. Rev. Lett.} \textbf{{97}}, {120405} ({2006}).

 \bibitem{Branciard2012}
 {{C.} {Branciard}},
 {{E. G.} {Cavalcanti}}, 
 {{S. P.} { Walborn}}, 
  {{V.} {Scarani}}, {and}
 {{H. M.} {Wiseman}},
{Phys. Rev. A} \textbf{{85}}, {010301} ({2012}).

\bibitem{bell}
 {{J.} {Bell}}, 
 {Phys.} \textbf{{1}},
 {195} ({1964}).

\bibitem{bell-review}
 {{N.} {Brunner}}, 
 {{D.} {Cavalcanti}},
 {{S.} {Pironio}},   
 {{V.} {Scarani}},   {and}
 {{S.} {Wehner}}, 
 {Rev. Mod. Phys.} \textbf{{86}},
 {419} ({2014}).

\bibitem{Wiseman2007}
 {{H. M.} {Wiseman}}, 
 {{S. J.} {Jones}},  {and}
 {{A. C.} {Doherty}}, 
 {Phys. Rev. Lett.} \textbf{{98}},
 {140402} ({2007}).

\bibitem{Cavalcanti2007}
 {{D.} {Cavalcanti}},  {and}
 {{P.} {Skrzypczyk}}, 
 {Rep. Prog. Phys.} \textbf{{80}},
 {024001} ({2017}).
 
 \bibitem{Uola2019}
 {{R.} {Uola}},
 {{A. C. S.} {Costa}},
 {{H. C.} {Nguyen}},  {and}
 {{O.} {G\"uhne}}, 
 {arXiv:1903.06663}.
 

 \bibitem{Bennet2012}
 {{A.J.} {Bennet}},
 {{D.A.} {Evans}},
 {{D.J.} {Saunders}},
 {{C.} {Branciard}},
 {{E.G.} {Cavalcanti}},
 {{H.M.} {Wiseman}}, {and}
 {{G.J.} {Pryde}}, 
 {Phys. Rev. X} \textbf{2}, {031003} ({2012}).
 
 \bibitem{Woodhead2016}
 {{E.} {Woodhead}}, 
 {New J. Phys.} \textbf{{18}}, ({2016}).
    
\bibitem{Weston2018}
	{{M. M.} {Weston}}, 
	{{S.} {Slussarenko}}, 
	{{H.M.} {Chrzanowski}}, 
	{{S.} {Wollmann}}, 
	{{L.K.} {Shalm}}, 
	{{V.B.} {Verma}}, 
	{{M.S.} {Allman}}, 
	{{S.W.} {Nam}}, {and} 
	{{G.J.} {Pryde}},
	{Sci. Adv.} \textbf{4}, {e1701230} ({2018}).
	
	 
 \bibitem{Wollmann2018}
 { {S.}  {Wollmann}}, 
 { {M. J. W.}  {Hall}}, 
 { {R. B.}  {Patel}},  
 { {H. M.}  {Wiseman}},  {and}
 { {G. J.}  {Pryde}}, 
   {Phys. Rev. A} \textbf{{98}},
   {022333} ({2018}).

\bibitem{Schneeloch13}
 { {J.}  {Schneeloch}}, 
 { {C. J.}  {Broadbent}}, 
 { {S. P.}  {Walborn}},  
 { {E. G.}  {Cavalcanti}},  {and}
 { {J. C.}  {Howell}}, 
   {Phys. Rev. A} \textbf{{87}},
   {062103} ({2013}).	 
	 
\bibitem{Costa2018}
 {{A. C. S.} {Costa}}, 
 {{R.} {Uola}},  {and}
 {{O.} {G\"uhne}}, 
 {Phys. Rev. A} \textbf{{98}},
 {050104(R)} ({2018}). 
 
 \bibitem{Krivachy2018}
 { {T.}  {Kriv\'achy}}, 
 { {F.}  {Fr\"owis}},  {and}
 { {N.}  {Brunner}}, 
   {Phys. Rev. A} \textbf{{98}},
   {062111} ({2018}).

\bibitem{footnote} 
{The results in Ref.~\cite{Schneeloch13} are special cases of Ref.~\cite{Costa2018} and~\cite{Krivachy2018}.} 

\bibitem{Moroder2016}
 {{T.} {Moroder}}, 
 {{O.} {Gittsovich}}, 
 {{M.} {Huber}}, 
 {{R.} {Uola}}, {and}
 {{O.} {G\"uhne}}, 
 {Phys. Rev. Lett.} \textbf{{116}},
 {090403} ({2016}). 

\bibitem{Shadboldt2012}
 {{P.} {Shadboldt}}, 
 {{T.} {Vertesi}}, 
 {{Y.-C.} {Liang}}, 
 {{C.} {Branciard}},
 {{N.} {Brunner}} {and}
 {{J. L.} {O'Brien}}, 
 {Sci. Rep.} \textbf{{2}},
 {420} ({2012}).
 
 \bibitem{Palsson2012}
 {{M.S.} {Palsson}}, 
 {{J.J.} {Wallman}}, 
 {{A.J.} {Bennet}}, {and}
 {{G. J.} {Pryde}}, 
 {Phys. Rev. A} \textbf{{86}},
 {032223} ({2012}).

 \bibitem{Havrda1967}
 { {J.}  {Havrda}}  {and}
 { {F.}  {Charvat}}, 
    {Kybernetika} \textbf{{3}},
   {30} ({1967}).

\bibitem{Tsallis1988}
 { {C.}  {Tsallis}}, 
   {J. Stat. Phys.} \textbf{{52}},
   {479} ({1988}).
   
\bibitem{Renyi1970}
 { {A.}  {R\'enyi}}, 
   {Valószínüségszámítás; Tankönyvkiadó: Budapest, Hungary (1966). (English Translation: Probability
Theory (North-Holland, Amsterdam, 1970)).} 
   
\bibitem{coverthomas}
 { {T. M.}  {Cover}}  {and}
 { {J. A.}  {Thomas}}, 
   {``Elements of Information Theory", Second edition, John Wiley \& Sons, 2006}. 

 \bibitem{Costa2018a}
 {{A. C. S.} {Costa}}, 
 {{R.} {Uola}},  {and}
 {{O.} {G\"uhne}}, 
 {Entropy} \textbf{{20}},
 {763} ({2018}).   
 
\bibitem{MU1988}
 { {H.}  {Maassen}},  {and}
 { {J.}  {Uffink}}, 
   {Phys. Rev. Lett.} \textbf{{60}},
   {1103} ({1988}).

\bibitem{Werner1989}
 { {R. F.}  {Werner}},
   {Phys. Rev. A} \textbf{{40}},
  {4277} ({1989}).

\bibitem{Rastegin2013}
 { {A. E.}  {Rastegin}}, 
   {Eur. Phys. J. D} \textbf{{67}},
   {269} ({2013}).

\bibitem{footnote2}
	{In Ref.~\cite{Krivachy2018}, the authors mention that these values of $r$ and $s$ perform optimally, which we have verified numerically in our framework.}

\bibitem{Kim2006}
 { {T.}  {Kim}}, 
 { {M.}  {Fiorentino}},  {and}
 { {F. N. C.}  {Wong}}, 
   {Phys. Rev. A} \textbf{{73}},
   {012316} ({2006}).

\bibitem{Fedrizzi2007}
 { {A.}  {Fedrizzi}}, 
 { {T.}  {Herbst}}, 
 { {A.}  {Poppe}},  
 { {T.}  {Jennewein}},  {and}
 { {A.}  {Zeilinger}}, 
   {Opt. Express} \textbf{{15}},
   {15377} ({2007}).

\bibitem{White2007}
 { {A. G.}  {White}}, 
 { {A.}  {Gilchrist}}, 
 { {G. J.}  {Pryde}},  
 { {J. L.}  {O’Brien}},
 { {M. J.}  {Bremner}},  {and}
 { {N. K.}  {Langford}}, 
   {J. Opt. Soc. Am. B} \textbf{{24}},
   {172} ({2007}).

\bibitem{Cavalcanti2015}
{{D.} {Cavalcanti}},
{{P.} {Skrzypczyk}},
{{G. H.} {Aguilar}},
{{R. V.}{Nery}},
{{P. H.} {Souto Ribeiro}}, {and}
{{S. P.}{Walborn}},
{Nat. Comm.} \textbf{{6}},
{7941} ({2015}). 

\bibitem{Riccardi2018}
 { {A.}  {Riccardi}}, 
 { {C.}  {Macchiavello}},  {and}
 { {L.}  {Maccone}}, 
   {Phys. Rev. A} \textbf{{97}},
   {052307} ({2018}).

\bibitem{Yang2019}
 { {H.}  {Yang}}, 
 { {Z.-Y.}  {Ding}},
 { {D.}  {Wang}},
 { {H.}  {Yuan}},
 { {X.-K.}  {Song}},
 { {J.}  {Yang}},
 { {C.-J.}  {Zhang}}, {and}
 { {L.}  {Ye}}, 
   {Phys. Rev. A} \textbf{{101}},
   {022324} ({2020}).
   

\bibitem{Zhao2019}
 { {Y.-Y.}  {Zhao}}, 
 { {C.}  {Zhang}}, 
 { {S.}  {Cheng}}, 
 { {X.}  {Li}}, 
 { {Y.}  {Guo}}, 
 { {B.-H.}  {Liu}}, 
 { {H.-Y.}  {Ku}}, 
 { {S.-L.}  {Chen}}, 
 { {Q.}  {Wen}}, 
 { {Y.-F.}  {Huang}}, 
 { {G.-Y.}  {Xiang}}, 
 { {C.-F.}  {Li}},  {and}
 { {G.-C.}  {Guo}}, 
   {arXiv:1909.13432}.



\end{thebibliography}
\end{document}